\documentclass[lettersize,journal]{IEEEtran}
\usepackage{amsmath,amsfonts}
\usepackage{algorithmic}
\usepackage{algorithm}
\usepackage{array}
\usepackage[caption=false,font=normalsize,labelfont=sf,textfont=sf]{subfig}
\usepackage{textcomp}
\usepackage{stfloats}
\usepackage{url}
\usepackage{verbatim}
\usepackage{graphicx}
\usepackage{cite}
\usepackage{color}
\usepackage{multirow}
\usepackage{booktabs}
\usepackage{hyperref}
\hypersetup{pdfborder={0 0 0}}
\usepackage[normalem]{ulem}
\usepackage{soul}
\usepackage{xcolor}

\usepackage{tcolorbox}

\begin{document}

\title{Task-Oriented Communication for Human Action Understanding via Edge-Cloud Co-Inference}

\author{Jingyi Liu,
        Cheng Yuan,
        Lijun He,~\IEEEmembership{Member,~IEEE},
        Jun Zhang,~\IEEEmembership{Fellow,~IEEE},
        and Jiawei Shao,~\IEEEmembership{Member,~IEEE}
    
\thanks{J. Liu, C. Yuan, and J. Shao are with the Institute of Artificial Intelligence of China Telecom (TeleAI), China (E-mail: jingyiliu@stu.xjtu.edu.cn, yuanc3@chinatelecom.cn, shaojw2@chinatelecom.cn). J. Liu and L. He are with the School of Information and Communications Engineering, Xi'an Jiaotong University, Xi'an, China (E-mail: lijunhe@mail.xjtu.edu.cn). J. Zhang is with the Department of Electronic and Computer Engineering, Hong Kong University of Science and Technology, Hong Kong (E-mail: eejzhang@ust.hk). The corresponding author is J. Shao.}
}

\maketitle

\begin{abstract}
The expanding application of smart sensing has created a growing demand for the accurate understanding of human action at the network edge. Traditional approaches require massive video data to be transmitted from resource-constrained edge devices to powerful cloud servers, incurring prohibitive uplink bandwidth consumption and unacceptable latency while raising privacy concerns. To overcome these bottlenecks, we propose a task-oriented communication framework for human action understanding (TOAU) through edge-cloud collaboration. Our framework utilizes a monocular pose estimator to extract continuous joint coordinates from raw videos, followed by a vector quantized variational autoencoder (VQ-VAE) to convert these coordinates into discrete motion tokens. Consequently, only a compact sequence of codebook indices is transmitted over the network, consuming as few as 2.25 bits per motion frame and avoiding privacy leakages. At the cloud server, a lightweight projector aligns these motion tokens with the embedding space of a large vision-language model (VLM) to facilitate complex action understanding, which is trained with an efficient instruction tuning paradigm. Comprehensive evaluations on three benchmarks demonstrate that TOAU reduces the transmission payload to approximately 1\% of that required by video codec-based solutions and achieves substantially lower end-to-end latency under bandwidth-constrained uplink conditions, while maintaining comparable action understanding accuracy.
\end{abstract}


\section{Introduction}
The proliferation of smart cities and ubiquitous camera networks \cite{smartcamera1, smartcamera3} has driven a growing demand for intelligent sensing, while the rapid development of autonomous systems further expands the scope of real-world applications. Many of these applications are inherently human-centered \cite{humancenter}, requiring accurate perception of human trajectory, actions, and intent. Representative examples include smart surveillance \cite{smartsurveillance}, elderly care \cite{elder}, and healthcare monitoring \cite{healthcare}, where understanding human motion is essential for behavior analysis and anomaly detection. Beyond passive monitoring scenarios, accurate action understanding also plays a fundamental role in autonomous robots, human–computer interaction, and embodied multi-agent systems. 
For instance, service robots \cite{servicerobot} rely on action understanding to provide context-aware assistance, while collaborative robots \cite{crobot2, crobot4} leverage motion perception to ensure effective cooperation, and embodied agents \cite{ea1, ea2} continuously interpret human behavior to enable adaptive responses and smooth human–agent interaction.
In these scenarios, the ability to accurately capture and interpret human motion in real time is the dominant factor in system reliability and user experience. 

Due to the development of these technologies, human action understanding has evolved into a key capability for next-generation intelligent systems. Apart from the ability to accurately align action sequences with textual descriptions, recent tasks \cite{motionllm, llamo, haic} increasingly involve more complex reasoning processes about human action understanding, such as spatial-temporal comprehension, intent prediction, and integration with action policy. These tasks require models to capture rich contextual information and perform cross-modal reasoning beyond low-level perception. To address these challenges, large foundation models have emerged as a promising solution, owing to their strong capability in semantic representation and multimodal reasoning. However, deploying such models on edge devices remains highly challenging due to their substantial computational and memory requirements. In most practical scenarios, resource-constrained edge devices cannot support large foundation models with billions of parameters. Even under optimized settings, the resulting inference latency remains prohibitive, rendering real-time processing infeasible. As a result, the dominant framework consists of transmitting visual content captured by edge devices to powerful cloud servers, where computationally intensive vision–language model (VLM) inference can be performed.

In practice, directly transmitting raw RGB video streams from edge devices to the cloud is sub-optimal for three main reasons. First, continuous video transmission from large-scale camera networks consumes substantial uplink bandwidth and leads to unacceptable end-to-end latency, especially in high-concurrency or mobile environments. Second, raw videos contain numerous pixel-level visual details, many of which are irrelevant to human action understanding. Background textures, illumination variations, and other environmental details contribute little to the target task, yet occupy the majority of data transmission. Third, uploading visual data inevitably exposes sensitive information about both the environment and individuals, raising critical privacy concerns in human-centered applications.

These inefficiencies originate from a fundamental mismatch between direct video transmission and the requirements of downstream machine tasks. Traditional video transmission methods are primarily designed to reconstruct the complete scene and preserve pixel-level details for human viewing, whereas human action understanding depends only on task-relevant information, which typically occupies a minor portion of the video content. To bridge this gap, a paradigm shift from conventional data-oriented transmission to task-oriented communication is required. Instead of transmitting full video streams captured by edge devices, task-oriented communication aims to extract and deliver only the minimal essential information necessary for a specific downstream task. In the context of human action understanding, the semantic essence of the task lies predominantly in human kinematic movements, while the accurate reconstruction of pixel-level visual content is largely redundant. 

Driven by this task-oriented philosophy, we develop an edge–cloud cooperative inference framework for human action understanding to minimize data transmission and system latency. Specifically, the edge device extracts human movements from input videos and encodes these motion features into compact representations, reducing the communication cost to as low as 2.25 bits per frame. Upon receiving the transmitted data, the cloud server recovers the motion features and performs high-level semantic reasoning with a pretrained large language model (LLM) with a trained motion projector. Moreover, the motion feature of each video frame constitutes only a single token during LLM inference, significantly fewer than mainstream VLMs that convert a high-definition frame into hundreds of visual tokens, thus reducing computational complexity in the cloud. By reformulating the transmission target from pixel-level details to action-centric representations, the proposed framework substantially reduces uplink communication overhead and end-to-end latency, while maintaining comparable performance in human action understanding tasks.

\begin{figure}[!t]
\centering
\includegraphics[width=0.9\columnwidth]{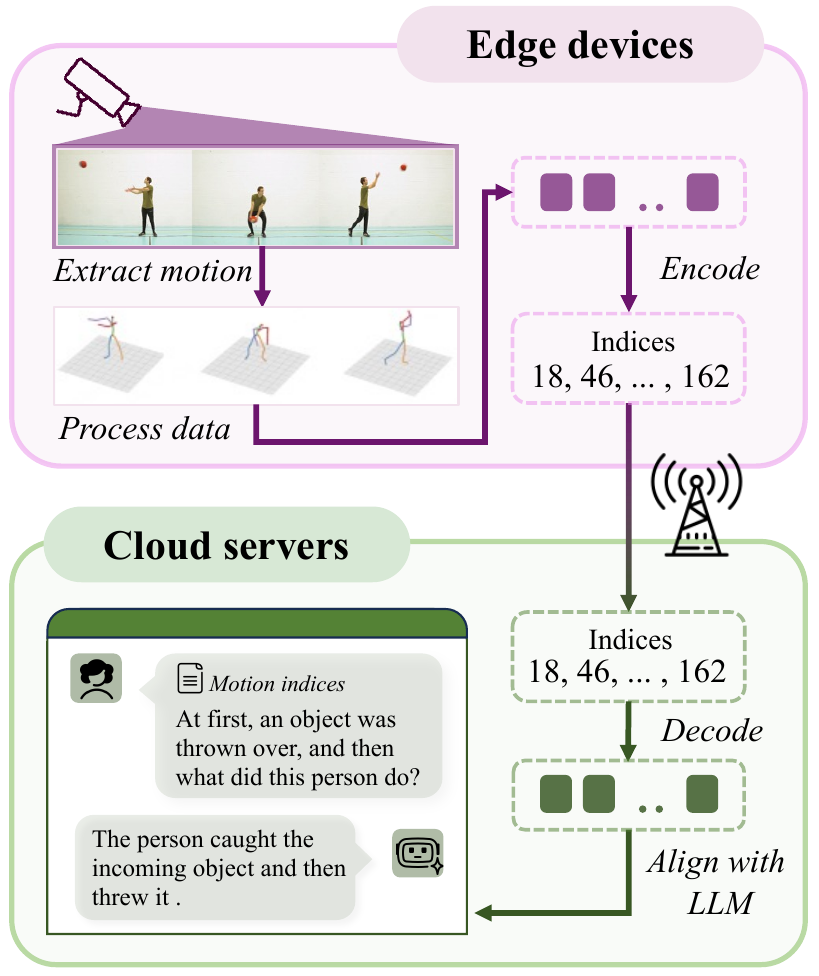}
\caption{An example of the proposed task-oriented action understanding system. The edge device extracts motion data from raw videos and converts them into compact indices, while the cloud server decodes motion representations and handles the intensive LLM reasoning to generate responses.}
\label{fig_1}
\end{figure}

\subsection{Contributions}
As illustrated in Fig.~\ref{fig_1}, we propose a task-oriented human action understanding system (TOAU) based on edge-cloud cooperative inference. In summary, the main contributions of this paper are outlined as follows: 
\begin{itemize}
\item \textbf{Task-Oriented Edge-Cloud Framework for Action Understanding:}
We propose an end-to-end edge-cloud collaborative system for human action understanding, where motion extraction is performed on edge devices and intensive reasoning is handled by the cloud. The proposed framework effectively decouples task-relevant information from the large volume of video data, enabling efficient transmission tailored to downstream inference. In addition, the proposed motion representation requires approximately one discrete motion token for every four motion frames, contrary to the hundreds of visual tokens produced by typical VLMs.
\item \textbf{Compact Motion Representation for Efficient Transmission:}
We employ an efficient representation for the extracted motion features using a vector quantized variational autoencoder (VQ-VAE), which transmits only compact indices in a codebook shared by both edge devices and cloud servers. This design reduces communication costs to as low as 2.25 bits per motion frame while preserving the essential information required for accurate action understanding.
\item \textbf{Comprehensive Evaluation of Transmission-Accuracy Trade-off:}
We conduct extensive experiments on three benchmarks to comprehensively evaluate the proposed TOAU system. Results demonstrate that TOAU achieves a favorable trade-off between transmission cost and task performance, reducing the transmission payload to approximately 1\% of that required by video-based approaches and the VLM input token overhead by approximately 50\%, while maintaining comparable action understanding accuracy and achieving lower end-to-end latency under bandwidth-constrained uplink conditions.

\end{itemize}

\subsection{Organization}
The rest of this paper is organized as follows. Section II provides a detailed description of related works. Section III introduces the system model, and Section IV elaborates on the proposed TOAU method. Experiment results are presented and analyzed in Section V, and Section VI concludes the paper.

\section{Related Works}

\subsection{Task-Oriented Communication}
The increasing demand for large-scale data transmission between user devices and cloud servers introduces significant burdens on the mobile network. Task-oriented communication has recently emerged as a new paradigm \cite{TOC2, TOC3, TEM} to minimize data transmission. The principle of this paradigm is to extract a compact representation of the information necessary to complete a specific downstream task. Unlike conventional data-oriented communication, which serves as a reliable bitstream pipe for raw input data, task-oriented communication removes irrelevant information and delivers only task-relevant information to reduce bandwidth consumption and transmission latency while maintaining task performance \cite{TOC4}. 

Early studies primarily investigated general frameworks for efficient task-oriented communication. Based on the information bottleneck (IB) principle, the VFE method \cite{shao2022edgeinference} formalizes the trade-off between communication overhead and inference accuracy for single-device edge inference. This framework was subsequently extended to multi-device edge systems through distributed coding to eliminate redundancy among different devices \cite{shao2023multidevice}.

More recent studies have explored task-oriented communication in increasingly specialized application scenarios. For edge video analytics, Shao et al. \cite{TEM} proposed a task-oriented framework that combines IB-based feature extraction with a temporal entropy model to exploit temporal correlations among video features, thereby reducing transmission overhead while supporting downstream inference. Other studies have applied task-oriented communication to cooperative perception in autonomous driving \cite{shao2024vehicle} and device-edge co-inference of large multimodal models \cite{TOFC}. Further developments have addressed robustness against domain shifts \cite{shao2025shift} and task-oriented transmission over digital communication systems \cite{xie2023robust}.

These studies highlight that the design of task-oriented communication systems should be closely aligned with the information requirements of specific downstream tasks. In particular, human action understanding relies heavily on human motion dynamics and temporal relationships, which motivates the use of motion-centric representations rather than transmitting complete visual information. Following this insight, we investigate task-oriented communication for human action understanding via edge-cloud co-inference, leveraging compact motion representations to reduce communication overhead while preserving task-relevant information.

\subsection{Neural Video Compression}
A wide range of video compression techniques have been utilized to reduce video transmission costs while preserving essential visual content. Conventional standardized video codecs remain dominant solutions in practical deployments, including H.264/AVC \cite{264}, H.265/HEVC \cite{265}, AV1 \cite{AV1}, and H.266/VVC \cite{VVC}. These standards integrate inter-frame prediction, motion estimation and compensation, transform coding, and entropy coding to optimize the rate–distortion (RD) performance. However, these codecs rely on hand-crafted modules and fail to fully exploit the spatio-temporal redundancy inherent in videos.

To overcome these limitations, learning-based video compression methods \cite{dvc, dcvc, dcvc2, dcvc3, dcvc4} have emerged as a promising alternative, which employ neural networks to extract compact representations of the video content and reconstruct frames from entropy-coded latent features. The encoding and decoding networks are trained in an end-to-end manner to learn the optimal transformation from a large volume of video data, thus enabling more flexible RD trade-offs. For instance, DVC \cite{dvc} builds an end-to-end learned video compression framework by replacing key modules in traditional codecs, including motion estimation, residual coding, and entropy modeling, with neural networks. In this way, the motion and residual information can be jointly optimized under the rate-distortion objective rather than being designed through hand-crafted rules. 
More recently, DCVC-UF \cite{dcvc4} introduces a chunk-based coding framework that jointly models spatiotemporal dependencies across multiple frames, enabling efficient compression and parallel reconstruction. By streamlining entropy coding, it further improves the trade-off between compression efficiency and computational complexity.
In parallel, generative compression approaches \cite{videocp1} have been introduced to synthesize perceptually plausible details and improve human perceptual quality for low bitrate transmission. 

The aforementioned approaches focus on reconstructing the entire video, which contains substantial task-irrelevant information such as background environment and human identity. In contrast, we reformulate the goal according to the downstream task and transmit only the human action in the video necessary for machine understanding, thus advancing the boundaries of RD optimization.

\subsection{3D Human Motion Estimation and Processing} 
In recent years, video-based methods for estimating human pose and shape \cite{vibe, motionv2, motionv3, motionv4} have shown greater practicality than single-image approaches \cite{e2e, hpsloop, cliff, singf1}, as they exploit temporal information for more stable predictions. Many existing works follow a standard pipeline: localize human subjects using object detectors, and then reconstruct 3D human motion sequences from monocular videos by predicting the parameters of the skinned multi-person linear (SMPL) model \cite{smpl}.

A representative work \cite{vibe} incorporates recurrent neural networks to model temporal dependencies across consecutive frames, significantly improving the stability of motion reconstruction. Nevertheless, monocular 3D human motion estimation remains inherently ill-posed due to the absence of explicit depth information, leading to ambiguities, especially under self-occlusion and complex motions. To alleviate this issue, a subsequent work \cite{cliff} introduces explicit camera parameter modeling \cite{geometric} together with full-image contextual features. This approach reduces depth ambiguity and enhances reconstruction accuracy by formulating motion estimation in a consistent global coordinate system. More recent works further improve temporal consistency and the ability to capture detailed motions. For example, PMCE \cite{pmce} directly estimates SMPL mesh parameters instead of intermediate joint representations, thus producing more natural and coherent motions. Another approach \cite{wham} leverages a pretrained encoder to construct synthetic 2D keypoints \cite{vitpose} and ground-truth motion sequence pairs to achieve more robust estimations.
Although multi-view systems can mitigate depth ambiguity through geometric triangulation, their reliance on multiple calibrated cameras limits practical deployment. Monocular video-based motion estimation therefore provides a more accessible approach to obtaining 3D skeleton-based motion from ordinary RGB videos.

\begin{figure*}[!t]
\centering
\includegraphics[width=\textwidth]{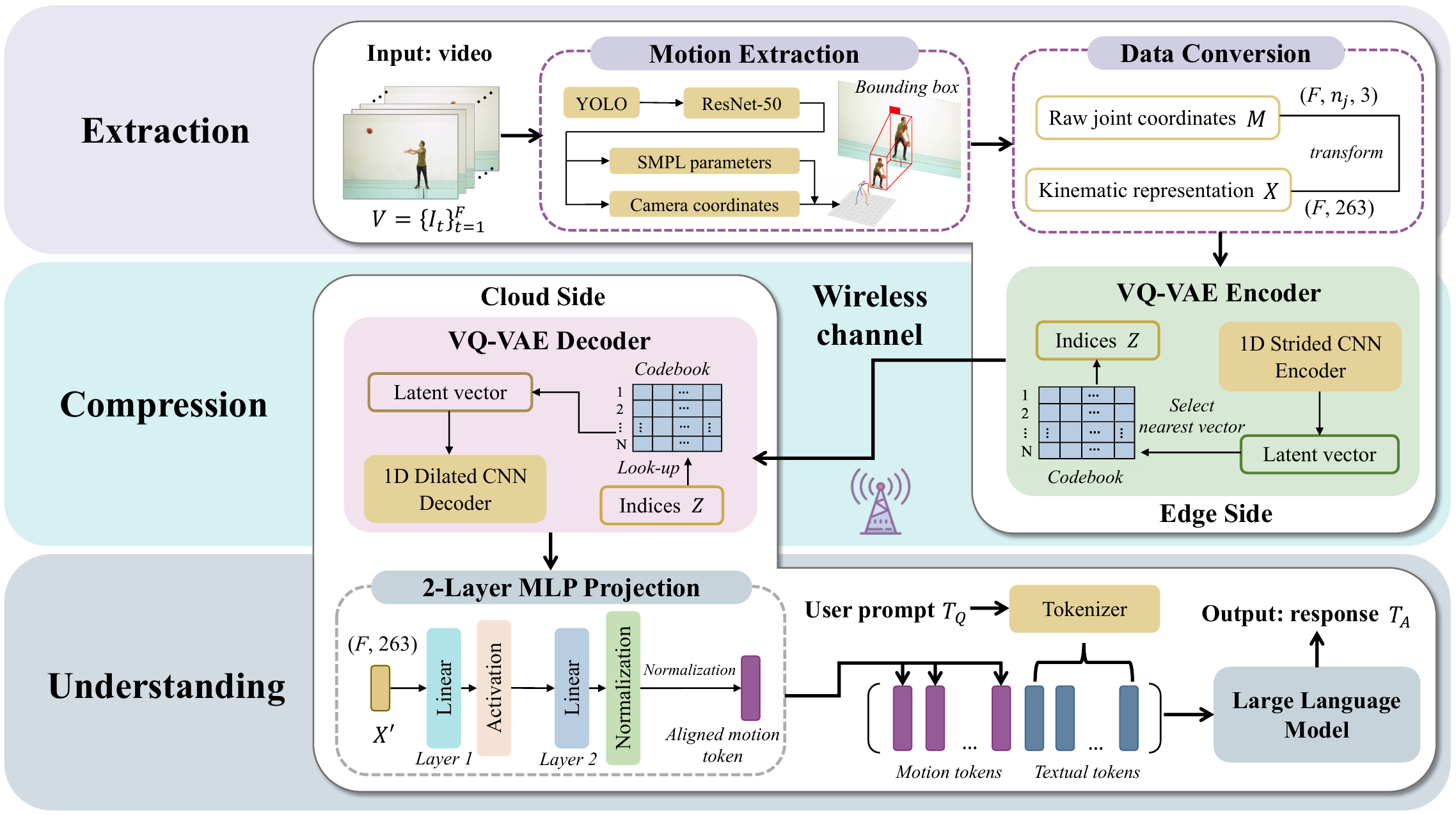}
\caption{System framework of the proposed action understanding pipeline for edge-cloud co-inference. The edge side performs monocular human pose estimation on input videos and transforms the results into motion representations. These representations are compressed into discrete VQ-VAE codebook indices. Four video frames only require as few as one index (9 bits) to be transmitted over the network, minimizing system latency and avoiding privacy leakages. The cloud side reconstructs the motion
representations and maps them into the embedding space of a pretrained LLM through a trained motion projector, enabling motion-conditioned reasoning. }
\label{fig_2}
\end{figure*}

Beyond 3D pose estimation, recent studies have further explored the effective utilization of structured skeletal motion representations. For instance, motion retargeting methods preserve motion
semantics across different skeletal structures \cite{motionretarg}, while skeleton-based action recognition approaches extract discriminative spatiotemporal features from joint sequences \cite{strongerskel}. Furthermore, informative sample selection strategies \cite{selecmodel} improve data efficiency by identifying valuable skeletal sequences for model training. These studies highlight the importance of structured motion information for downstream tasks, motivating our exploration of compact motion representations for task-oriented communication and human action understanding.

\subsection{Human Action Understanding}

The goal of human action understanding is to extract semantic information about human activities from multimodal data, encompassing fine-grained motion captioning, behavior analysis, and intention reasoning. Early research primarily focuses on bridging human motion and natural language through structured or statistical representations. For instance, this method \cite{statistical} modeled the relationship between motion primitives and textual descriptions using statistical methods. Under predefined syntactic rules, PoseScript \cite{posescript} further introduced a pipeline that converts low-level pose features into structured textual descriptions. Similarly, this work \cite{translation} treated motion as a special language and employed paired recurrent autoencoders for bidirectional translation between motion sequences and descriptions.

Subsequent works have explored representation learning to bridge motion and language and enable effective motion understanding. Inspired by vision–language pre-training, an early approach \cite{motionclip} mapped continuous 3D motion sequences into the contrastive language-image pre-training (CLIP) latent space to align them with textual descriptions. While such alignment provides a strong semantic foundation, it does not support a direct understanding of motion. Another line of work models motion as discrete token sequences similar to lingual tokens and leverages sequence modeling to capture their relationships. This study \cite{tm2t} departs from explicit mapping strategies by representing motion as discrete tokens and learning their temporal and semantic relationships through sequence modeling. More recent frameworks \cite{avatargpt, moitongpt} further extend this paradigm by integrating motion representations into LLMs for enhanced reasoning.

More recently, the emergence of LLMs and VLMs has motivated researchers to treat motion understanding as an independent task, in contrast to previous paradigms that integrate understanding and generation. MotionLLM \cite{motionllm} pioneers this paradigm \cite{llava, videollava} by employing a projector for motion representations and conducting instruction tuning on motion-text data, thereby empowering the model \cite{vicuna} to comprehend action nuances based on natural language instructions. Building upon this paradigm, LLaMo \cite{llamo} incorporates cross-attention mechanisms for instruction-guided semantic integration. These methods collectively highlight that human motion should not be treated merely as auxiliary visual information, but rather as a standalone input modality. 

\section{System Model}
As illustrated in Fig.~\ref{fig_2}, our proposed task-oriented communication framework for human action understanding conducts cooperative inference between resource-constrained edge devices and a powerful cloud server. Unlike traditional paradigms that rely on sending raw video streams, our framework extracts motion data from input videos and transmits only compact, task-relevant representations. As a result, the edge device is solely responsible for lightweight human pose estimation and motion compression, while the cloud server undertakes multimodal alignment and LLM reasoning.

Given a monocular RGB video $V = \{I_t\}^F_{t=1}$ with $F$ frames, the edge device first employs a human pose estimator to extract a continuous 3D motion sequence $M = \{m_t\}^F_{t=1}$. Initially, this raw motion sequence is represented as a spatial-temporal tensor $M \in \mathbb{R}^{F \times n_{\rm j} \times 3}$, with each element denoting the estimated world coordinate of a specific joint in a video frame. To bridge the gap between raw joint coordinates and high-level action semantics, we transform $M$ into the HumanML3D format, yielding a highly structured motion representation $X \in \mathbb{R}^{F \times 263}$. This data representation directly incorporates spatial-temporal motion dynamics such as joint velocities and rotations. To further reduce the communication overhead, $X$ is fed into a VQ-VAE encoder, which performs temporal downsampling by a factor of $l$ and quantizes the motion features into discrete codebook indices.
This process yields an index sequence $I = \{i_t\}^{F'}_{t=1}, i_t\in \{0, \dots, K-1\}$, where $K$ denotes the codebook size. Therefore, only the 1D sequence $I$ is required to be transmitted over the wireless channel. 
With each index consuming only $\log_2 K$ bits, the total transmission cost is merely $F' \log_2 K$ bits, forming an ultra-compact task-oriented communication payload.

After receiving the motion index sequence $I$, the cloud server performs semantic alignment and task-oriented reasoning. Specifically, the transmitted indices are first decoded through the shared VQ-VAE codebook to reconstruct the motion representation $X' \in \mathbb{R}^{F \times 263}$. The recovered features are then processed by a lightweight motion projector $P_\theta$, which transforms $X'$ into a sequence of motion embeddings $E \in \mathbb{R}^{F \times d_e}$, where $d_e$ is the embedding dimension of the language model. $P_{\theta}$ bridges the modality gap between the motion sequence $X'$ and linguistic embeddings by aligning motion features with the semantic space of the LLM. In this way, the aligned motion embeddings $E$ not only become semantically meaningful to the foundation model, but also make it possible to exploit the rich prior knowledge within the LLM. Finally, based on the user's query $T_{\rm Q}$ in natural language, the model reasons on the aligned motion embeddings to generate a fine-grained response $T_{\rm A}$, without requiring cloud-side video reconstruction.

In this work, we consider a bandwidth-constrained uplink with reliable delivery of transmitted motion indices. The effects of transmission errors and packet losses are not explicitly modeled, while error-control mechanisms and their associated overheads are left for future investigation.

\section{Methods}

\subsection{Task-Oriented Motion Extraction at the Edge}
We deploy a human pose estimation module on edge devices to extract the 3D human joint coordinates from the raw video. Given a raw video frame $I_t \in \mathbb{R}^{H \times W \times 3}$, an off-the-shelf object detector first localizes the human subject to obtain a bounding box $b_t$, which is further fed into the model to provide spatial context for global rotation estimation. The estimator then takes both the cropped human region and the corresponding bounding box $b_t$ to regress the body pose parameters (e.g., the joint rotations) $\theta_t \in \mathbb{R}^{72}$, the body shape parameters $\beta_t \in \mathbb{R}^{10}$, and the global camera parameters. Subsequently, the skinned multi-person linear (SMPL) mesh deformation function $\mathcal{M}(\cdot)$ is employed on the body parameters to obtain the 3D joint locations $J_t \in \mathbb{R}^{n_{\rm j} \times 3}$, written as:
\begin{equation}
\label{eq1}
J_t = \mathcal{M}(\theta_t, \beta_t).
\end{equation}
The obtained 3D joints $J_t$ are defined in a root-centered coordinate system. 

To recover the pose in the 3D camera coordinate system associated with the original full image, a global translation is applied, yielding the full-image 3D joints.
The global translation is calculated from the predicted crop-based weak-perspective camera parameters $(s_t,t_{x,t},t_{y,t})$:

{
\begin{equation}
\label{eq:camera_translation}
\begin{aligned}
t_{x,t}^{\mathrm{full}}
&= t_{x,t} + \frac{2(c_{x,t}-W/2)}
{b_t^{\mathrm{crop}}s_t},\\
t_{y,t}^{\mathrm{full}}
&= t_{y,t} + \frac{2(c_{y,t}-H/2)}
{b_t^{\mathrm{crop}}s_t},\\
t_{z,t}^{\mathrm{full}}
&= \frac{2f}{b_t^{\mathrm{crop}}s_t}.
\end{aligned}
\end{equation}
}

Here, $(c_{x,t},c_{y,t})$ denotes the bounding-box center in the original image, and $b_t^{\mathrm{crop}}=200\,\mathrm{scale}_t$ is the effective crop size after bounding-box adjustment. The principal point is assumed to be $(W/2,H/2)$, and the focal length is estimated as $f=\sqrt{W^2+H^2}$ following CLIFF. This transformation compensates for the crop offset and converts the predicted camera parameters into the full-image camera reference under the assumed intrinsics:

\begin{equation}
\label{eq2}
J_t^{\rm full} = J_t + \mathbf{1}(\mathbf{t}_t^{\rm full})^\top,
\end{equation}
where $t^{\rm full} \in \mathbb{R}^{3}$ denotes the global translation, and $\mathbf{1}$ is an all-ones column vector used to broadcast the translation to all joints. Through this mechanism, the system effectively distills the high-volume RGB video into a sequence of 3D spatial coordinates $J \in \mathbb{R}^{F \times n_{\rm j} \times 3}$, removing visual appearance information such as background details and textures.

While the recoverd 3D joint coordinates represent the skeletal structure, they are highly sensitive to camera translations and body scales. To facilitate robust semantic alignment for the cloud-side LLM, we transform these raw coordinates into a viewpoint-invariant kinematic representation $X \in \mathbb{R}^{F \times 263}$ in the HumanML3D \cite{humanml3d} format. Specifically, the physical pose $p$ in each motion frame is structured as a tuple with eight components, given by:
\begin{equation}
\label{eq3}
p = (\dot{r}^a, \dot{r}^x, \dot{r}^z, r^y, j^p, j^r, j^v, c^f).
\end{equation}
Here, $\dot{r}^a \in \mathbb{R}$ denotes the root angular velocity about the Y-axis, $(\dot{r}^x,\dot{r}^z) \in \mathbb{R}^2$ represent the root linear velocities in the XZ-plane, and $r^y \in \mathbb{R}$ denotes the root height \cite{rotation}. The variables $\mathbf{j}^p \in \mathbb{R}^{3 \times (n_{\rm j}-1)}$ and $\mathbf{j}^r \in \mathbb{R}^{6 \times (n_{\rm j}-1)}$ represent the root-relative joint positions and continuous 6D joint rotations, respectively, excluding the root joint. Meanwhile, $\mathbf{j}^v \in \mathbb{R}^{3 \times n_{\rm j}}$ represents the local joint velocities, and $\mathbf{c}^f \in \mathbb{R}^4$ denotes the foot-contact features. The root-related features characterize the overall movement trajectory and orientation changes, while the local joint positions, rotations, and velocities capture fine-grained variations in body posture and motion.

\subsection{Discrete Tokenization for Ultra-Low Data Transmission}
Even though the edge device extracts the kinematic representation $X \in \mathbb{R}^{F \times 263}$ from the video, transmitting such dense floating-point tensors remains prohibitive for bandwidth-constrained edge networks. To achieve extreme data compression while preserving the underlying human action semantics, we utilize the encoder of a VQ-VAE on the edge device to quantize the continuous motion representation into a finite set of discrete tokens.

Specifically, the continuous motion sequence $X$ is first processed by a 1D convolutional encoder $\mathcal{E}(\cdot)$ to produce a latent representation:
\begin{equation}
Z_e = \mathcal{E}(X).
\end{equation}
Taking advantage of the temporal redundancy inherent in human motion, the encoder simultaneously projects the motion feature to a latent dimension $d$ and performs temporal downsampling with a factor of $l$. This operation yields a continuous latent representation $Z_e \in \mathbb{R}^{F' \times d}$, where $F' = \lfloor F/l \rfloor$ and $\lfloor \cdot \rfloor$ denotes the rounding down operation.

To discretize these latent representations for efficient transmission over the network, we construct a finite learnable codebook:
\begin{equation}
\mathcal{C} = \{c_0, c_1, \dots, c_{K-1}\} \in \mathbb{R}^{K \times d},
\end{equation}
with a vocabulary size of $K$. Each temporal slice $z_{e,t}$ of the latent vector $Z_e$ is converted into its nearest codebook entry via the nearest-neighbor quantization. The quantized latent vectors can be represented by a sequence of discrete indices $I = \{i_1, i_2, \dots, i_{F'}\}$, mathematically formulated as:
\begin{equation}
\label{minimum-vae}
i_t = \arg\min_{k \in \{0, 1, \dots, K-1\}} \| z_{e,t} - c_k \|_2.
\end{equation}
Consequently, the continuous motion information is encapsulated in this sequence of integers, consuming only a total of $F' \lceil \log_2 K \rceil$ bits, where $\lceil \cdot \rceil$ denotes the rounding up operation.

Upon receiving these indices, the cloud server first retrieves the corresponding vectors $Z_q \in \mathbb{R}^{F' \times d}$ from the shared codebook $\mathcal{C}$.
Subsequently, the VQ-VAE decoder reverts the temporal downsampling and reconstructs the HumanML3D kinematic representation, denoted as $X' \in \mathbb{R}^{lF' \times 263}$.

This discrete tokenization pipeline transforms the edge-cloud communication paradigm. Instead of transmitting visually complete scenes or dense latent features, our task-oriented framework exclusively transmits the compact index sequence $I$ to achieve minimal bandwidth consumption and transmission latency. 

\subsection{Multimodal Alignment and Reasoning at the Cloud}
We adopt the widely recognized visual instruction tuning paradigm \cite{llava} to efficiently train the LLM deployed at the cloud to understand and reason on motion representations. By bridging the gap between the kinematic latent space and the LLM embedding space, the cloud server translates pure motion data into high-level semantics relevant to human action understanding tasks.

Since the backbone model cannot understand this form of motion data, we design a lightweight motion projector $P_\theta$ to align the reconstructed motion data $X'$ with the embedding space of the LLM. The resulting motion tokens $E \in \mathbb{R}^{lF' \times d_{\rm e}}$ are computed as:
\begin{equation}
\label{mlp}
E_{\rm motion} = P_\theta(X'),
\end{equation}
where $P_\theta(\cdot)$ denotes a learnable projection module that maps the reconstructed motion data into the embedding space of the LLM.

To integrate the kinematic representation $E$ with the natural language context, we implement an efficient placeholder substitution strategy. First, we define a special token, denoted as \texttt{<motion>}, to signify a motion token corresponding to a single frame. For a motion sequence consisting of $lF'$ frames, we initialize a unified discrete token sequence $T_{\rm seq}$ by concatenating the structural separators, the \texttt{<motion>} placeholders repeated $lF'$ times, the user prompt $T_{\rm Q}$, and the ground-truth answer $T_{\rm A}$:
\begin{equation}
\begin{split}
\label{T-seq}
T_{\rm seq} = [&\texttt{<start>}, \underbrace{\texttt{<motion>}, \dots, \texttt{<motion>}}_{lF' \text{ frames}},\\  
&\texttt{<end>}, T_{\rm Q}, T_{\rm A}].
\end{split}
\end{equation}
Here, $T_{\rm A}$ is only available during training and is omitted at inference time, where the model generates the answer autoregressively conditioned on the preceding tokens. This discrete sequence is then mapped to the embedding space of the foundation model. For the natural language components, namely the question $T_{\rm Q}$ and the answer $T_{\rm A}$, the corresponding embeddings in the LLM's pretrained vocabulary matrix are selected to constitute the textual embedding sequence $E_{\rm text}$. In contrast, the embedding positions for \texttt{<motion>} placeholders are dynamically replaced by our projected motion tokens $E_{\rm motion}$. Consequently, the final multimodal input tensor $S_{\rm in}$, which is fed into the LLM's Transformer blocks, is formulated as the concatenation of these continuous representations:
\begin{equation}
S_{\rm in} = [e_\texttt{<start>} \oplus E_{\rm motion} \oplus e_\texttt{<end>} \oplus E_{\rm text}],
\end{equation}
where $\oplus$ denotes the concatenation operation along the sequence dimension, and $e_{(\cdot)}$ represents the textual embedding of the corresponding special token. With a custom placeholder token introduced to the backbone model's context, this substitution mechanism integrates the kinematic features into the original feature space without altering the pretrained linguistic embeddings.

Given the multimodal input $S_{\rm in}$, the model is optimized using a standard autoregressive cross-entropy objective.
The training data is constructed in an instruction-tuning format, where each sample consists of a motion sequence, a user query $T_{\rm Q}$, and a target response $T_{\rm A}$. This formulation enables the model to align motion representations with natural language supervision for task-oriented reasoning.
Leveraging the strong prior knowledge embedded in the pretrained VLM, our approach requires only a small amount of task-specific data for effective adaptation, resulting in low training cost and efficient fine-tuning.

\section{Experiment}
\subsection{Experiment Setup}
\subsubsection{Network Implementation}
On the edge side, we first employ YOLO26 \cite{yolo26} to detect and localize human subjects in monocular videos, followed by the CLIFF model \cite{cliff} for 3D human joint coordinates estimation.
Specifically, the 49 joints obtained from CLIFF are converted into a skeletal representation with $n_{\rm j}=22$ joints following the KIT skeleton template, using joint selection and interpolation as needed.
Following the preprocessing pipeline of HumanML3D \cite{humanml3d}, the resulting $(22,3)$ joint coordinates are transformed into a 263-dimensional kinematic representation. To enable efficient communication, we adopt the VQ-VAE model of T2M-GPT \cite{t2mgpt} for motion compression.
The encoder performs temporal downsampling with a factor of $l = 4$, reducing the temporal length of the motion sequence from $F$ to $F'=\lfloor F/l \rfloor$. With a codebook size of $K=512$, each latent feature is quantized into a discrete codebook index, which can be represented using only 9 bits.
With $K=512$ and $l=4$, each discrete motion index requires 9 bits, corresponding to a nominal coding cost of 2.25 bits per motion frame for task-relevant motion information. 
On the cloud side, we employ the LLM component of Qwen3-VL-8B \cite{qwen3vl} as the backbone model and introduce a lightweight two-layer MLP as the motion projector, which maps motion features into the embedding space of the LLM to enable motion-conditioned reasoning. 

In the following experiments, edge-side computation is evaluated on a desktop workstation equipped with an NVIDIA GeForce RTX 4060 GPU (8 GB) and an Intel Core i5-12400F CPU, while cloud-side inference is performed on an NVIDIA H100 GPU (80 GB HBM3).
 
\subsubsection{Training Details}
The primary objective of the motion projector training is to establish a semantic mapping between continuous motion representations and the native embedding space of the LLM. To this end, we adopt a progressive training strategy using three motion–language datasets with complementary characteristics. Specifically, HumanML3D provides single-sentence descriptions for motion capture sequences derived from the AMASS dataset \cite{amass}, offering stable and noise-free data for coarse motion–language alignment. The training set of BABEL-QA \cite{babelqa} further introduces structured supervision on fine-grained motion attributes, such as body parts and motion directions. In addition, Motion-X \cite{motionx} is constructed from real-world videos. It enriches the data sources with more diverse and detailed action descriptions, thus improving robustness in realistic scenarios.

During training, the LLM backbone is kept frozen to preserve its language understanding and reasoning capabilities. The model is optimized using a standard autoregressive cross-entropy objective, which is applied solely to update the motion projector $P_\theta$. By treating motion descriptions as the target sequence $Y$, the projector learns to align kinematic representations with high-level action semantics in the language space.

\begin{table*}[!t]
    \caption{Evaluation Results on Motion-Bench}
    \centering
    \label{tab1}
    \renewcommand{\arraystretch}{1.2}
    \makebox[\textwidth][c]{%
    \begin{tabular}{|c|c|c|c|c|c|c|c|c|c|c|}
    \hline
    \textbf{Method} & \textbf{Codec} & \textbf{Turn acc.} & \textbf{Intent acc.} & \textbf{Event acc.} & \textbf{Order acc.} & \textbf{Movement acc.} & \textbf{Total acc.} & \textbf{Size (KB)} & \textbf{\# tokens}\\
    \hline
    
    \multirow{4}{*}{\shortstack[c]{Qwen3-VL\\ (320$\times$180)}} 
    & H.264 
    & 26.7 & 37.5 & 51.4 & 57.1 & 52.5 & 47.5
    & 21.45 & \multirow{4}{*}{1156.20} \\
    \cline{2-9}
    
    & H.265 
    & 23.3 & 43.8 & 50.0 & 57.1 & 49.2 & 46.0
    & 15.58 &  \\
    \cline{2-9}
    
    & AV1 
    & 23.3 & 37.5 & 47.2 & 57.1 & 47.5 & 44.0
    & 13.06 &  \\
    \cline{2-9}
    
    &DCVC-UF (LD)
    & 23.3 & 43.8 & 45.8 & 57.1 & 45.9 & 43.5
    & 9.98 &  \\
    \hline
    
    \multirow{4}{*}{\shortstack[c]{Qwen3-VL\\ (160$\times$90)}} 
    & H.264 
    & 36.7 & 37.5 & 43.1 & 52.4 & 42.6 & 42.5
    & 12.76 & \multirow{4}{*}{289.05} \\
    \cline{2-9}
    
    & H.265 
    & 40.0 & 43.8 & 44.4 & 52.4 & 45.9 & 45.0
    & 10.91 &  \\
    \cline{2-9}
    
    & AV1 
    & 36.7 & 37.5 & 41.7 & 57.1 & 47.5 & 44.0
    & 8.30 &  \\
    \cline{2-9}
    
    &DCVC-UF (LD)
    & 40.0 & 43.8 & 44.4 & 42.8 & 47.5 & 44.5
    & 5.00 &  \\
    \hline
    
    \multirow{4}{*}{\shortstack[c]{Qwen3-VL\\ (32$\times$32)}} 
    & H.264 
    & 40.0 & 37.5 & 41.7 & 42.9 & 41.0 & 41.0 
    & 6.59 & \multirow{4}{*}{19.27} \\
    \cline{2-9}
    
    & H.265 
    & 36.7 & 50.0 & 41.7 & 42.9 & 42.6 & 42.0
    & 7.17&  \\
    \cline{2-9}
    
    & AV1 
    & 40.0 & 43.8 & 41.7 & 42.9 & 39.3 & 41.0
    & 4.80 &  \\
    \cline{2-9}

    &DCVC-UF (LD)
    & 36.7 & 43.8 & 38.9 & 33.3 & 45.9 & 40.5
    &1.51 &  \\
    \hline

    \multicolumn{2}{|c|}{Ours}
    & 53.3 & 25.0 & 45.8 & 38.1 & 59.0 & 48.5 & 0.04 & 136.96 \\
    \hline

    \end{tabular}%
    }
\end{table*}

\begin{table*}[!t]
\caption{Evaluation Results on MoVid-Bench}
\centering
\label{tab2}
\renewcommand{\arraystretch}{1.2}

\makebox[\textwidth][c]{%
\begin{tabular}{|c|c|c|c|c|c|c|c|c|c|c|c|c|c|c|c|}

\hline
\multirow{2}{*}{\textbf{Method}} 
& \multirow{2}{*}{\textbf{Codec}}
& \multicolumn{2}{c|}{\textbf{Body}} 
& \multicolumn{2}{c|}{\textbf{Direction}} 
& \multicolumn{2}{c|}{\textbf{Hallucination}} 
& \multicolumn{2}{c|}{\textbf{Reasoning}} 
& \multicolumn{2}{c|}{\textbf{Sequence}} 
& \multicolumn{2}{c|}{\textbf{Total}} 
& \multirow{2}{*}{\shortstack[c]{\textbf{Size}\\ \textbf{(KB)}}}
& \multirow{2}{*}{\textbf{\# tokens}} \\
\cline{3-14}

& 
& \textbf{Acc.} & \textbf{Score} 
& \textbf{Acc.} & \textbf{Score} 
& \textbf{Acc.} & \textbf{Score} 
& \textbf{Acc.} & \textbf{Score} 
& \textbf{Acc.} & \textbf{Score} 
& \textbf{Acc.} & \textbf{Score} 
&  &  \\
\hline

\multirow{4}{*}{\shortstack[c]{Qwen3-VL\\ (160$\times$90)}} 
& H.264 
& 39.5 & 2.63 
& 40.0 & 2.40 
& 12.5 & 1.88 
& 67.9 & 3.43 
& 24.0 & 1.87 
& 40.6 & 2.52 
& 4.86 & \multirow{4}{*}{213.75} \\
\cline{2-15}

& H.265 
& 34.9 & 2.77 
& 40.0 & 2.50 
& 12.5 & 1.50 
& 66.1 & 3.41 
& 20.0 & 1.91 
& 37.5 & 2.55 
& 7.44 &  \\
\cline{2-15}

& AV1 
& 37.2 & 2.86 
& 40.0 & 3.40 
& 37.5 & 2.25 
& 69.7 & 3.68 
& 41.3 & 2.61 
& 48.4 & 3.01 
& 7.27 &  \\
\cline{2-15}

&DCVC-UF (LD)
& 46.5 & 3.21
& 40.0 & 3.10
& 25.0 & 1.88
& 73.2 & 3.75
& 40.0 & 2.67
& 50.5 & 3.09
& 4.51 & \\
\hline

\multirow{4}{*}{\shortstack[c]{Qwen3-VL\\ (80$\times$46)}} 
& H.264 
& 27.9 & 2.07
& 40.0 & 2.60
& 12.5 & 1.25
& 60.7 & 3.32 
& 25.3 & 1.93
& 36.5 & 2.38
& 6.94 & \multirow{4}{*}{28.50} \\
\cline{2-15}

& H.265 
& 27.9 & 2.37 
& 30.0 & 2.50 
& 12.5 & 1.12 
& 62.5 & 3.38 
& 26.7 & 2.07 
& 37.0 & 2.50 
& 7.45 &  \\
\cline{2-15}

& AV1 
& 30.2 & 2.26 
& 40.0 & 2.60 
& 12.5 & 1.12 
& 62.5 & 3.45 
& 26.7 & 1.96 
& 38.0 & 2.46 
& 6.88 &  \\
\cline{2-15}

&DCVC-UF (LD)
& 32.6 & 2.21
& 40.0 & 2.60
& 12.5 & 1.63
& 64.3 & 3.46
& 28.0 & 2.07
& 39.6 & 2.52
& 3.80 & \\
\hline

\multirow{4}{*}{\shortstack[c]{Qwen3-VL\\ (32$\times$32)}} 
& H.264 
& 25.6 & 1.88 
& 30.0 & 2.50 
& 12.5 & 1.50 
& 55.4 & 3.07 
& 20.0 & 1.69 
& 31.8 & 2.17 
& 5.27 & \multirow{4}{*}{14.25} \\
\cline{2-15}

& H.265 
& 20.9 & 1.74 
& 30.0 & 2.50 
& 12.5 & 1.62 
& 55.4 & 3.07 
& 25.3 & 1.83 
& 32.8 & 2.20 
& 6.51 &  \\
\cline{2-15}

& AV1 
& 27.9 & 1.98 
& 30.0 & 2.50 
& 12.5 & 1.50 
& 55.4 & 3.07 
& 20.0 & 1.51 
& 32.3 & 2.12 
& 4.22 &  \\
\cline{2-15}

&DCVC-UF (LD)
& 20.9 & 1.84
& 30.0 & 2.50
& 12.5 & 1.38
& 62.5 & 3.21
& 21.3 & 1.64
& 33.3 & 2.18
& 1.45 & \\
\hline

\multicolumn{2}{|c|}{Ours}
& 39.5 & 2.91 
& 10.0 & 1.80 
& 62.5 & 3.50 
& 67.9 & 3.73 
& 29.3 & 2.29 
& 43.2 & 2.88 
& 0.03 & 105.69 \\
\hline

\end{tabular}%
}
\end{table*}

\subsubsection{Evaluation Details}
The action understanding capability is evaluated on the following three benchmarks.

BABEL-QA is constructed based on the BABEL \cite{babel} and AMASS datasets. In BABEL, long motion sequences are first segmented into short clips, each annotated with a textual description, and the final action annotations are obtained by concatenating these segment-level captions. The benchmark includes questions regarding action categories, motion directions, and temporal movements of body parts. Due to the structured construction of this benchmark, both its questions and answers follow relatively rigid patterns. Therefore, we convert the original open-ended questions into a multiple-choice format by enumerating all possible answers within each category.

Built upon real-world videos, MoVid-Bench \cite{motionllm} evaluates action understanding and reasoning capabilities from multiple perspectives. Formulated as a free-form question answering task, this benchmark contains scenario-specific questions with more diverse and detailed answers. For evaluation, we adopt the official prompt templates provided by the authors and utilize Qwen3-VL-32B as the judge model for both accuracy and scores.

Unlike existing benchmarks that may rely heavily on textual cues inherent in question prompts or suffer from inaccurate alignment between machine-generated annotations and actual motions \cite{motionllm}, Motion-Bench is designed to emphasize motion-dependent reasoning. Specifically, we build the benchmark based on the AMASS dataset, ensuring accurate and fine-grained motion representations. We manually curate a subset of motion sequences and annotate question–answer pairs with a focus on minimizing information leakage from instructions, thus facilitating accurate measurements on action understanding capabilities.

Motion-Bench is designed to comprehensively evaluate five different aspects of motion understanding, including:
\begin{itemize}
    \item \textbf{Turn}: to identify rotational movements and their frequency.
    \item \textbf{Event}: to recognize the occurrence of specific actions.
    \item \textbf{Intent}: to infer the underlying purpose of actions.
    \item \textbf{Movement}: to estimate spatial movements in the physical world.
    \item \textbf{Order}: to understand the temporal ordering of actions.
\end{itemize}

Among all categories, the Intent questions are specifically designed for more complex motion reasoning. To alleviate the influence of prior knowledge and linguistic biases, we avoid questions whose answers can be easily inferred without observing the motion. Instead, we formulate questions that require understanding the underlying purpose behind human actions. For example, rather than asking about body parts involved in an action, we instruct the model to infer why the legs are bent during a motion sequence. This type of question requires a holistic analysis of human action in addition to accurate motion recognition. 

\subsection{Result Analysis}

\subsubsection{Rate-Performance Trade-off}
In practical scenarios, the uplink transmission bandwidth of the user device is often constrained, and thus input videos are typically compressed to different extents to meet communication requirements. We analyze the overall rate–performance trade-off by changing the video compression settings and measuring the resulting accuracy on human action understanding tasks. 
We evaluate two benchmarks under different video coding configurations by varying the video resolution, codec, and codec-specific compression parameters, including constant rate factor (CRF) for conventional codecs and the quantization parameters for I-frames ($QP_I$) and P-frames ($QP_P$) for DCVC-UF (LD).
For each setting, we report category-wise accuracy and score, overall accuracy, average transmission size, and average visual token count. The results on Motion-Bench and MoVid-Bench are presented in Table~\ref{tab1} and Table~\ref{tab2}, respectively. 

Taking DCVC-UF (LD) as a representative baseline for video compression, we conduct a detailed quantitative comparison. On our Motion-Bench, compared to DCVC-UF (LD) with a resolution of 160$\times$90, TOAU reduces the transmission size to only 0.04 kilobytes (KB), corresponding to 0.8\% of the compressed video size (5.00 KB), while achieving a 9.0\% relative improvement in accuracy over the video-based pipeline. Meanwhile, the average number of visual tokens decreases from 289.05 to 136.96, thus reducing the computational costs on the cloud. 
On MoVid-Bench, under the same video configuration, TOAU similarly achieves substantial reductions in transmission overhead, reducing the transmission size to just 0.7\% of the video baseline (4.51 KB). In addition, the token count is reduced by 51\%, further improving inference efficiency. However, unlike the results on Motion-Bench, TOAU exhibits an accuracy decrease from 50.5\% to 43.2\%, corresponding to a gap of 7.3 percentage points. This result highlights the trade-off between communication efficiency and task accuracy, suggesting that the compact motion representation may discard some visual information relevant to downstream action understanding.

These observations demonstrate the potential of task-oriented motion transmission to substantially reduce communication overhead compared with video codecs, while also revealing benchmark-dependent differences in downstream task accuracy. In particular, the stronger performance of DCVC-UF (LD) on MoVid-Bench suggests that visual information beyond human motion can be beneficial for certain action understanding tasks.

\begin{figure*}[!t]
\centering
\subfloat[]{\includegraphics[width=0.5\textwidth]{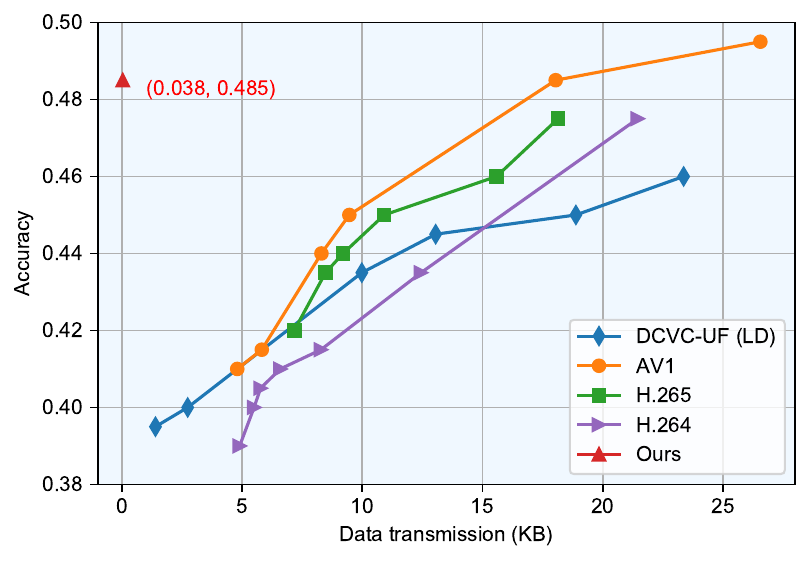}%
\label{result-motionQA}}
\hfil
\subfloat[]{\includegraphics[width=0.5\textwidth]{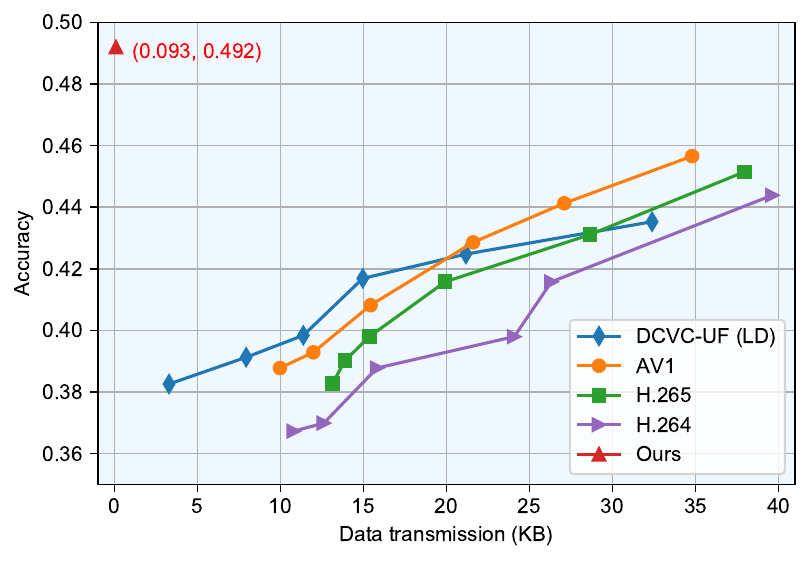}%
\label{result-babelQA}}
\caption{The rate-performance curves of different methods in action understanding benchmarks. Subfigures (a) and (b) show results on Motion-Bench and BABEL-QA, respectively. The proposed task-oriented communication system reduces the transmission payload by a factor of more than 30 while maintaining competitive overall accuracies in both benchmarks.}
\label{fig_result}
\end{figure*}

\begin{figure*}[!t]
\centering
\subfloat[]{\includegraphics[width=0.5\textwidth]{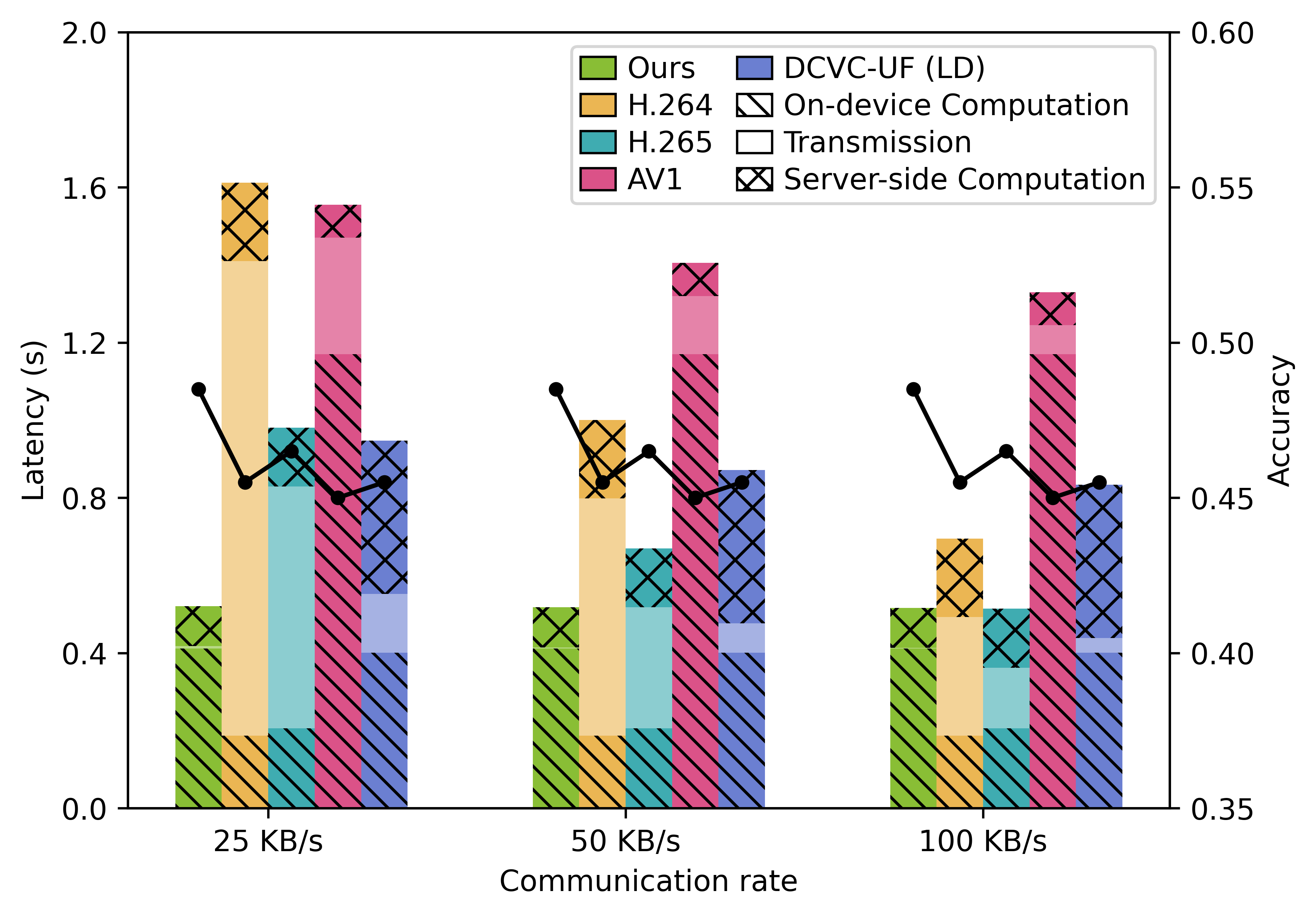}%
\label{time-motionQA}}
\hfil
\subfloat[]{\includegraphics[width=0.5\textwidth]{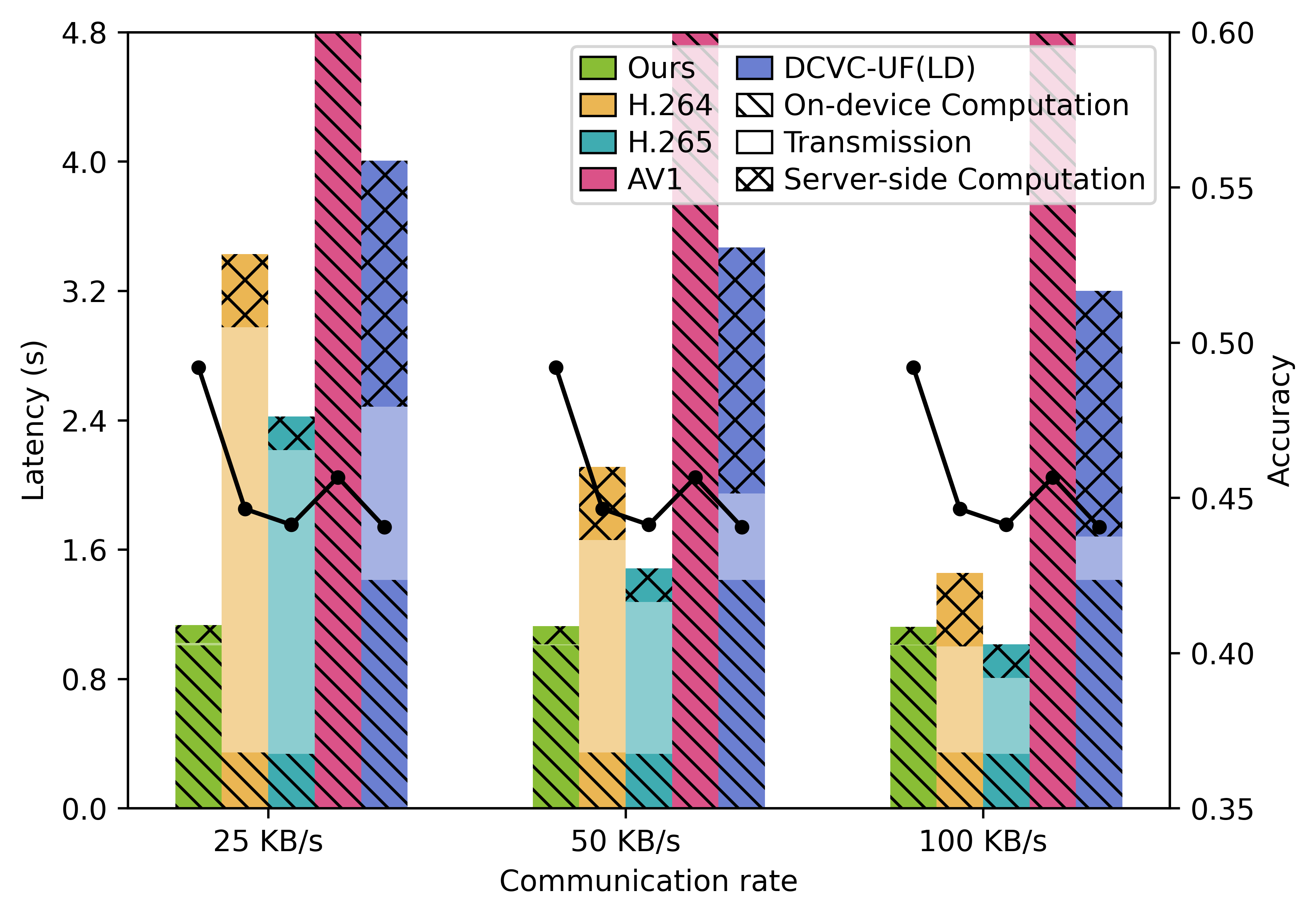}%
\label{time-babel}}
\caption{Average inference latency per user request (the bar plot corresponding to the left y-axis) and benchmark score (the line plot corresponding to the right y-axis) under different communication rates. The total latency consists of on-device computation time, data transmission latency, and server-side computation time. Subfigures (a) and (b) show results on Motion-Bench and BABEL-QA, respectively.}
\label{fig_time}
\end{figure*}

To further illustrate this relationship, Figs.~\ref{result-motionQA} and \ref{result-babelQA} plot accuracy as a function of transmission size on the proposed benchmark and BABEL-QA, respectively. The results show that our approach achieves similar accuracy with compressed videos with at least two orders of magnitude smaller data size. Among the video compression baselines, DCVC-UF (LD) achieves the lowest transmission sizes while maintaining competitive accuracy in the low-bitrate regime, demonstrating the compression efficiency of learning-based video codecs. However, as the transmission size increases, its task accuracy does not consistently outperform conventional codecs such as H.265 and AV1. This suggests that improved video compression efficiency does not necessarily translate into better preservation of task-relevant motion information.
These results demonstrate that pixel-based compression suffers from an undesirable trade-off: lowering video quality reduces transmission cost but leads to significant degradations in downstream accuracy for the task of human action understanding. In contrast, our method maintains strong performance under extremely constrained communication budgets by directly encoding task-relevant motion dynamics.

The category-wise results also reveal certain limitations of motion-only representations. 
Intent-related questions in Table~\ref{tab1} require inferring high-level action semantics from temporal motion patterns, which may not be fully captured or interpreted through motion representation. 
For Order, temporal compression may weaken fine-grained motion details, while overlapping answer choices further increase the difficulty of discrimination. Regarding Direction in Table~\ref{tab2}, some reference answers require scene-related information unavailable in skeleton-based motion data, and the small number of questions makes the accuracy sensitive to individual errors. These observations highlight the limitations of motion-only representations and the varying information requirements of different human action understanding tasks.

\begin{figure*}[!t]
\centering
\includegraphics[width=0.98\textwidth]{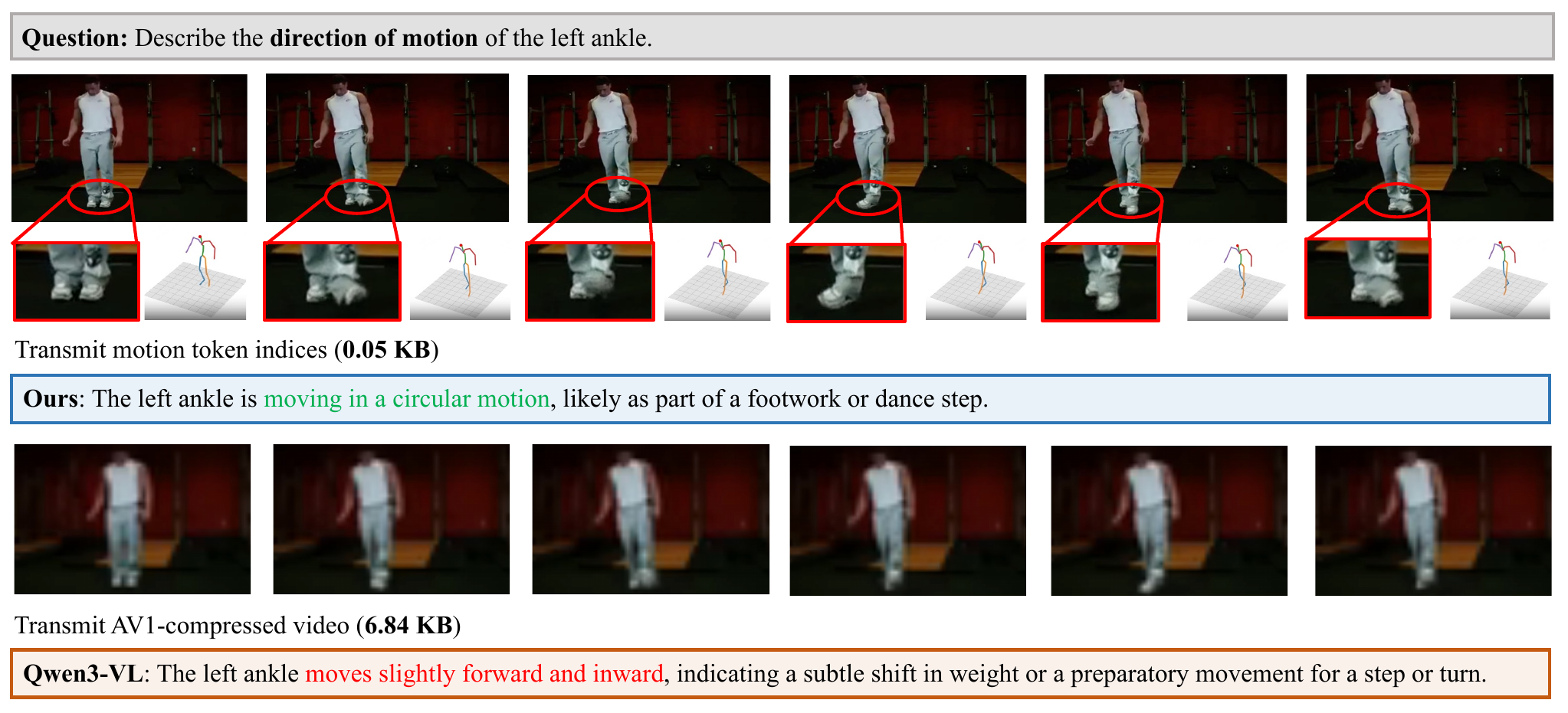}
\caption{A qualitative comparison between the proposed TOAU method and traditional AV1 compression. Under a low data rate, the AV1-compressed video is heavily blurred, rendering ankle movements difficult to identify. In contrast, the proposed method retains human motion details necessary for accurate action understanding.}
\label{fig_qa1}
\end{figure*}

\begin{figure}[!t]
\centering
\includegraphics[width=0.97\columnwidth]{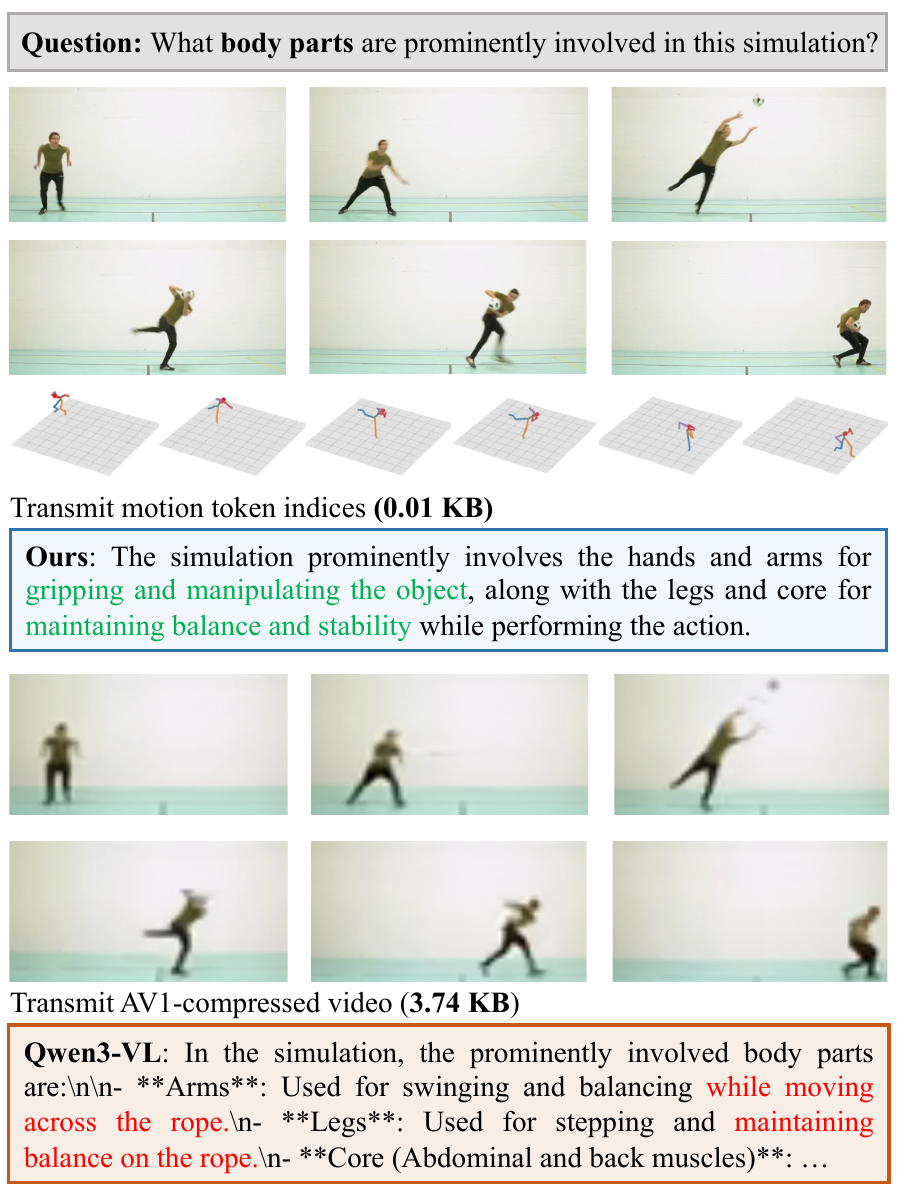}
\caption{A qualitative comparison between the proposed TOAU method and traditional AV1 compression. The motion semantics in the AV1 video are heavily distorted under a low data rate, and thus Qwen3-VL fails to infer the true intent behind this action sequence. In contrast, the proposed method preserves motion semantics and avoids hallucinations.}
\label{fig_qa2}
\end{figure}

\subsubsection{Latency Analysis}
We evaluate the system latency under different transmission strategies and communication bandwidths. Specifically, we compare the proposed motion-based representation with conventional video compression codecs H.264, H.265, and AV1, as well as the neural video codec DCVC-UF (LD). The conventional video codecs are implemented using FFmpeg with libx264 and libx265 (medium preset), and libaom-av1 (cpu-used = 4). The end-to-end system latency consists of three components: on-device computation, data transmission, and server-side computation. To demonstrate the impact of network conditions, we conduct experiments under multiple uplink bandwidth settings of 25, 50, and 100 KB/s.

As shown in Figs.~\ref{time-motionQA} and \ref{time-babel}, TOAU achieves the lowest end-to-end latency under bandwidth constraints of 25 and 50 KB/s on both Motion-Bench and BABEL-QA, while maintaining higher task accuracy than the video compression baselines. At 25 KB/s, TOAU reduces the total latency by 45.5\% compared with the fastest DCVC-UF (LD) baseline on
Motion-Bench, and by 53.7\% compared with H.265 on BABEL-QA. At 50 KB/s, the latency reductions remain 23.0\% and 24.5\%, respectively.
However, at 100 KB/s, H.265 achieves comparable latency to TOAU on Motion-Bench
(0.514 s versus 0.515 s) and lower latency on BABEL-QA (1.014 s versus 1.119 s). This behavior reflects the computation--communication trade-off: although TOAU incurs additional on-device computation for motion extraction, its compact motion representations substantially reduce transmission overhead, making it particularly advantageous under bandwidth-constrained conditions.

To provide a detailed comparison, we further analyze the latency of different stages. 
On Motion-Bench, TOAU incurs higher on-device computation latency (0.412 s) than conventional codecs such as H.264 (0.187 s), mainly due to the additional motion extraction process. Nevertheless, its transmission size is only 1.0\% of that of the most bitrate-efficient
video codec, DCVC-UF (LD). Moreover, TOAU achieves the lowest server-side computation latency (0.103 s), compared with 0.115 s for AV1 and 0.395 s for DCVC-UF (LD).

A similar trend is observed on BABEL-QA. Although TOAU incurs higher on-device latency (1.007 s) than H.264 (0.344 s) and H.265 (0.336 s), it remains faster than DCVC-UF (LD) (1.355 s) and AV1 (5.001 s). Meanwhile, its transmission size is only 0.4\% of that of the most bitrate-efficient video baseline (AV1). On the server side, TOAU achieves the lowest computation latency of 0.111 s, requiring only 53.2\% of the computation time of the fastest video baseline (H.265). We note that the generally higher computation latency on BABEL-QA is partly attributed to its longer video sequences, which increase the workload of motion extraction, video encoding/decoding, and downstream inference. Furthermore, despite the improved compression capability of the neural video codec DCVC-UF (LD), its computational overhead contributes to higher end-to-end latency on both benchmarks.
These results demonstrate that TOAU achieves a favorable computation--communication trade-off, substantially reducing end-to-end latency under bandwidth-constrained conditions while maintaining competitive human action understanding performance.

\begin{table*}[!t]
\caption{Ablation Study}
\centering
\label{tab3}
\renewcommand{\arraystretch}{1.2}
\makebox[\textwidth][c]{%
\begin{tabular}{|c|c|c|c|c|c|c|c|c|c|c|}

\hline
\multirow{2}{*}{\textbf{Method}} 
& \multirow{2}{*}{\textbf{Codec}}
& \multicolumn{3}{c|}{\textbf{BABEL-QA}} 
& \multicolumn{3}{c|}{\textbf{Motion-Bench}} 
& \multicolumn{3}{c|}{\textbf{MoVid-Bench}} \\
\cline{3-11}

& 
& \textbf{\# tokens} & \textbf{Size (KB)} & \textbf{Accuracy}
& \textbf{\# tokens} & \textbf{Size (KB)} & \textbf{Accuracy}
& \textbf{\# tokens} & \textbf{Size (KB)} & \textbf{Accuracy} \\
\hline

\multirow{4}{*}{\shortstack[c]{Qwen3-VL}} 
& H.264 
& \multirow{4}{*}{681.48} & 24.08 & 39.8
& \multirow{4}{*}{289.05} & 12.76 & 42.5
& \multirow{4}{*}{28.50}  & 6.94 & 36.5 \\
\cline{2-2} \cline{4-5} \cline{7-8} \cline{10-11}

& H.265 
& & 15.36 & 39.8
& & 10.08 & 43.0
& & 6.87 & 37.5 \\
\cline{2-2} \cline{4-5} \cline{7-8} \cline{10-11}

& AV1 
& & 15.43 & 40.8
& & 9.55 & 44.5
& & 5.49 & 38.0 \\
\cline{2-2} \cline{4-5} \cline{7-8} \cline{10-11}

& DCVC-UF (LD)
& & 7.33 & 39.3
& & 3.94 & 44.5
& & 4.56 & 38.5 \\
\hline

\multirow{4}{*}{\shortstack[c]{Qwen3-VL\\ (fine-tuned)}} 
& H.264 
& \multirow{4}{*}{681.48} & 17.69 & 41.6 
& \multirow{4}{*}{289.05} & 12.76 & 44.0
& \multirow{4}{*}{28.50}  & 6.94 & 37.5 \\
\cline{2-2} \cline{4-5} \cline{7-8} \cline{10-11}

& H.265 
&  & 15.60 & 43.4
&  & 10.91 & 44.5
&  & 7.45 & 37.0 \\
\cline{2-2} \cline{4-5} \cline{7-8} \cline{10-11}

& AV1 
& & 14.06 & 40.7
& & 9.46 & 44.5
& & 5.49 & 40.1 \\
\cline{2-2} \cline{4-5} \cline{7-8} \cline{10-11}

& DCVC-UF (LD)
& & 7.33 & 41.2
& & 3.94 & 45.0
& & 4.56 & 41.1 \\
\hline

\multicolumn{2}{|c|}{Ours}
& 340.55 & 0.09 & 49.2
& 136.96 & 0.04 & 48.5
& 105.69 & 0.03 & 43.2\\
\hline

\end{tabular}%
}
\end{table*}

It should be noted that the current TOAU implementation operates at a fixed coding rate determined by the pretrained VQ-VAE configuration. Unlike conventional video codecs with adjustable rate control, TOAU cannot dynamically adapt its coding rate to varying uplink bandwidths without adopting an alternative or retrained VQ-VAE. Under the considered communication model, varying the uplink bandwidth only affects the transmission latency, while the encoded motion payload remains unchanged. Due to its extremely compact representation, the transmission latency of TOAU contributes only a small fraction of the overall system latency, making its end-to-end latency relatively insensitive to bandwidth changes.

However, as the available bandwidth increases, the transmission advantage of TOAU gradually diminishes, while its additional edge-side motion extraction overhead becomes more significant. As shown in Fig.~\ref{time-babel}, TOAU exhibits higher end-to-end latency than H.265 at 100 KB/s. Therefore, TOAU is particularly advantageous under bandwidth-constrained conditions, whereas further optimization of edge-side motion estimation would be beneficial for reducing its overall system latency.

\subsubsection{Case Study}
The key advantage of our proposed TOAU method lies in its ability to perform fine-grained human action perception and understanding based solely on compact motion representations. To demonstrate this capability, we present two qualitative examples selected from the MoVid-Bench dataset. Under stringent bandwidth and latency constraints, the selected video samples are compressed using the proposed method and the traditional AV1 codec.

As illustrated in Fig.~\ref{fig_qa1}, when questioned about the direction of motion of the ankle, TOAU correctly identifies a circular movement pattern in the video (i.e., the foot rotates around the ankle joint) even at an ultra-low transmission cost of 0.05 KB. In contrast, Qwen3-VL with AV1-compressed video input (6.84 KB) produces inaccurate descriptions such as "moves forward and inward," which also compromises the subsequent intent prediction. Even when provided with higher-resolution video input (45.28 KB), the model still fails to capture the correct motion pattern and instead recognizes an "up and down" movement.

The user query in Fig.~\ref{fig_qa2} focuses on identifying relevant body parts and inferring the underlying intent of the action. When data transmission is minimized to 0.01 KB, our method still successfully predicts that the arm movement is associated with catching an object and recognizes that the action of legs and core serves to maintain balance. However, with an AV1-compressed video of 3.74 KB, Qwen3-VL enumerates the motion trajectories of every body part and hallucinates about the intention of the motion as moving across a rope.
These examples demonstrate that video compression heavily distorts task-relevant information for human action understanding under limited bandwidth conditions, whereas motion representations enable accurate perception and reliable reasoning.

\subsection{Ablation Study}
To evaluate the effectiveness of the proposed motion-based representation, we conduct an ablation study where the backbone VLM is directly fine-tuned with our human motion dataset. Specifically, we render the motion data used to train our motion projector into RGB videos and use these videos to fine-tune the Qwen3-VL-8B backbone with identical training setups. We compare three settings under the same backbone: (1) the original model without any fine-tuning, (2) the model fine-tuned with rendered videos, and (3) the model equipped with the proposed motion projector. All models are evaluated on three benchmarks under bandwidth-constrained settings. 

As shown in Table~\ref{tab3}, our method achieves a significantly better trade-off between communication cost and performance, compared with video-based finetuning. 
Specifically, the transmitted data sizes of our pipeline are reduced to 1.2\%, 1.0\%, and 0.7\% of the most efficient DCVC-UF (LD) video compression on BABEL-QA, Motion-Bench, and MoVid-Bench, while improving the accuracy by 13.4\%, 7.8\%, and 5.1\% over the best results among the four video codec baselines. 
These results demonstrate that, compared to pixel-based video compression, motion representations effectively eliminate redundant visual information while preserving task-relevant dynamics. The proposed motion projector further provides high-fidelity motion semantics aligned with the LLM embedding space, leading to more efficient and accurate action understanding.

\section{Conclusions}
In this paper, we proposed a task-oriented edge-cloud collaborative framework for human action understanding, aiming to reduce the communication overhead and system latency in bandwidth-constrained edge networks. Instead of transmitting compressed videos, the proposed framework extracts human motion sequences from raw videos at the edge and converts them into compact discrete motion tokens through vector quantization. As a result, only indices are transmitted to the cloud, requiring as few as 2.25 bits per motion frame while preserving user privacy by removing background and identity information. On the cloud side, a lightweight motion projector aligns the decoded motion representations with the embedding space of the LLM, enabling fine-grained action understanding and complex reasoning through efficient instruction tuning.
Experiments on three benchmarks demonstrate that the proposed framework substantially reduces communication overhead while maintaining competitive action understanding capability and achieving lower end-to-end latency under bandwidth-constrained uplink conditions.
Nevertheless, the latency advantage of TOAU is dependent on the available uplink bandwidth. As the bandwidth increases, the benefit of reduced transmission overhead gradually diminishes, while the additional computational overhead of edge-side motion extraction becomes increasingly significant. Consequently, TOAU may exhibit higher end-to-end latency than conventional video codecs under higher-bandwidth conditions. Further optimizing edge-side motion estimation and processing could help alleviate this limitation and improve the overall system efficiency.

These results highlight the potential of task-oriented communication as an efficient paradigm for deploying multimodal intelligence in future edge-cloud systems. Future research directions include the adoption of task-oriented edge-cloud co-inference systems in autonomous robots to facilitate complex and responsive action planning.

\bibliographystyle{IEEEtran}  
\bibliography{intro, motion, task, video}

\end{document}